\documentclass[twocolumn,showpacs]{revtex4}
\usepackage{amssymb}
\usepackage[dvips]{graphicx}

\begin{document}

\title{Theory of High Temperature Superconductivity in Doped Polar Insulators%
}
\author{A. S. Alexandrov}
\affiliation{Department of Physics, Loughborough University, Loughborough LE11 3TU,
United Kingdom\\
}

\begin{abstract}
Many high-temperature superconductors are highly polarizable ionic lattices
where the Fr\"ohlich electron-phonon interaction (EPI) with longitudinal
optical phonons creates an effective attraction of doped carriers virtually
equal to their Coulomb repulsion. The general multi-polaron theory is given
with both interactions being strong compared with the carrier kinetic energy
so that the conventional BCS-Eliashberg approximation is inapplicable. The
many-electron system is described by the polaronic t-J$_p$  Hamiltonian with
reduced hopping integral, $t$, allowed double on-site occupancy, large
phonon-induced antiferromagnetic exchange, $J_p\gg t$, and a
high-temperature superconducting state of small superlight bipolarons
protected from clustering.
\end{abstract}

\maketitle

\section{Introduction}
It seems plausible that the true origin of high-temperature
superconductivity is found in a proper combination of the finite-range
Coulomb repulsion with a significant finite-range EPI as suggested by a
growing number of experimental and theoretical studies \cite{aledev}. In
highly polarizable ionic lattices like cuprate superconductors both
interactions are quite strong (of the order of 1 eV) compared with the low
Fermi energy of doped carriers because of a poor screening  by non- or
near-adiabatic carriers \cite{alebra}. In those conditions the
BCS-Eliashberg theory \cite{me} breaks down because of the polaronic
collapse of the electron bandwidth \cite{ale0}.

The many-body theory for polarons has been developed for extremely weak and
strong EPI. In the weak-coupling limit this problem is reduced to the study
of a structure factor of the uniform large polaron gas \cite{LDB1977}. For
strong coupling the problem is reduced to on-site \cite{aleran} or
inter-site \cite{ale96,inter} small bipolarons on a lattice. A strong enhancement
of T$_c$ was predicted in the crossover region from the BCS-like polaronic
to BEC-like bipolaronic superconductivity due to a sharp increase of the
density of states in a narrow polaronic band \cite{ale0}, which is missing
in the so-called \emph{negative} Hubbard $U$ model. Nevertheless the theory
of dense polaronic systems in the intermediate coupling regime remains
highly cumbersome, in particular, when EPI competes with strong electron
correlations. Corresponding microscopic models with the on-site Hubbard
repulsion and the short-range Holstein EPI have been studied using powerful
numerical techniques \cite{trugman,bonca}.

In most analytical and numerical studies mentioned above and many others
both interactions are introduced as input parameters not directly related to
the material. Quantitative calculations of the interaction matrix elements
can be performed from pseudopotentials using the density functional theory
(DFT) \cite{bauer}. On the other hand, one can express the bare Coulomb
repulsion and EPI through material parameters rather than computing them
from first principles in many physically important cases \cite{mahan}. In
particular, for a polar coupling to longitudinal optical phonons (the Fr\"{o}%
hlich EPI), which is the major EPI in polar crystals, both the momentum
dependence of the matrix element, $M(\mathbf{q})$, and its magnitude are
well known, $|M(\mathbf{q})|=\gamma (q))\hbar \omega _{0}/\sqrt{2N}$ with a
dimensionless $\gamma (q)=\sqrt{4\pi e^{2}/\kappa \Omega \hbar \omega
_{0}q^{2}}$, where $\Omega $ is a unit cell volume, $N$ is the number of
unit cells in a crystal, $\omega _{0}$ is the optical phonon frequency, and $%
\kappa =\epsilon _{\infty }\epsilon _{0}/(\epsilon _{0}-\epsilon _{\infty })$%
. The high-frequency, $\epsilon _{\infty }$ and the static, $\epsilon _{0}$
dielectric constants are both measurable in a parent polar insulator. As is
well known, a two-particle bound state exists even in the weak-coupling
regime, $\lambda <0.5$, due to a quantum (exchange) interaction between two
large polarons forming a \emph{\ large bipolaron} \cite{aledev}($\lambda $
is the familiar EPI  constant of the BCS-Eliashberg theory). These weakly
coupled large pairs overlap in dense systems, so that their many-particle
ground state is a BCS-like superconductor with Cooper pairs (see below).

Here the analytical multi-polaron theory is given in the
strong-coupling regime for highly polarizable lattices with
$\epsilon_0 \gg 1$.

\section{Generic Hamiltonian and its canonical transformation}The
dielectric response function of strongly correlated electrons is \emph{%
apriori} unknown. Hence one has to start with a generic Hamiltonian
including \emph{unscreened} Coulomb and Fr\"ohlich interactions operating on
the same scale since any ad-hoc assumption on their range and relative
magnitude might fail,
\begin{eqnarray}
H &=-&\sum_{i,j} (T_{ij}\delta_{ss^\prime}+\mu \delta _{ij}) c_{i}^{\dagger
}c_{j} +{\frac{1}{{2}}}\sum_{i\neq j}{\frac{e^2}{{\epsilon_\infty |\mathbf{%
m-n}|}}}\hat{n}_{i}\hat{n}_j+\cr && \sum_{\mathbf{q},i}\hbar \omega _{0}\hat{%
n}_{i}\left[ u(\mathbf{m,q} ) d_{\mathbf{q} }+H.c.\right]+H_{ph}.
\label{hamiltonian}
\end{eqnarray}
Here $T_{ij}\equiv T(\mathbf{m-n})$ is the bare hopping integral, $\mu$ is
the chemical potential, $i=\mathbf{m},s$ and $j=\mathbf{n},s^{\prime }$
include both site $(\mathbf{m,n})$ and spin $(s,s^{\prime })$ states, $u(%
\mathbf{m,q})= (2N)^{-1/2}\gamma(q)\exp(i \mathbf{q \cdot m})$, $c_{i}, d_{%
\mathbf{q} }$ are electron and phonon operators, respectively, $\hat{n}%
_{i}=c^\dagger_i c_i$ is a site occupation operator, and $H_{ph}=\sum_{%
\mathbf{q}}\hbar \omega _{0}(d_{\mathbf{q} }^{\dagger }d_{\mathbf{q}}+1/2)$
is the polar vibration energy.

In highly polarizable lattices with $\epsilon _{0}\rightarrow \infty
$ the familiar Lang-Firsov (LF) \cite{fir} canonical transformation
$e^{S}$ is
particulary instrumental with $S=-\sum_{\mathbf{q},i}\hat{n}_{i}\left[ u(%
\mathbf{m,q})d_{\mathbf{q}}-H.c.\right] $. It shifts the ions to new
equilibrium positions changing the phonon vacuum, and removes most of \emph{%
both }interactions from the transformed Hamiltonian, $\tilde{H}=e^{S}He^{-S}$%
,
\begin{equation}
\tilde{H}=-\sum_{i,j}(\hat{\sigma}_{ij}\delta _{ss^{\prime }}+\tilde{\mu}%
\delta _{ij})c_{i}^{\dagger }c_{j}+H_{ph},  \label{trans}
\end{equation}%
where $\hat{\sigma}_{ij}=T({\mathbf{m-n}})\hat{X}_{i}^{\dagger }\hat{X}_{j}$
is the renormalised hopping integral involving the multi-phonon transitions
described with $\hat{X}_{i}=\exp \left[ \sum_{\mathbf{q}}u(\mathbf{m,q})d_{%
\mathbf{q}}-H.c.\right] $, and $\tilde{\mu}=\mu +E_{p}$ is the chemical
potential shifted by the polaron level shift,
\begin{equation}
E_{p}={\frac{2\pi e^{2}}{{\kappa }}}\int_{BZ}{\frac{d^{3}q}{{(2\pi )^{3}q^{2}%
}}}.  \label{shift}
\end{equation}%
Here, the integration goes over the Brillouin zone (BZ) and
$E_{p}=0.647$ eV in La$_{2}$CuO$_{4}$ \cite{alebra}. The
electron-phonon coupling constant is defined as $\lambda
=2E_{p}N(0)$. In the case of 2D carriers with a
constant bare density of states, $N(0)=ma^{2}/2\pi \hbar ^{2}$ per spin, Eq.(%
\ref{shift}) places cuprates in the strong-coupling regime, $\lambda \gtrsim
0.5$, if the bare band mass $m>m_{e}$ (here $a$ is the in-plane lattice
constant).

\begin{figure}[tbp]
\begin{center}
\includegraphics[angle=-00,width=0.45\textwidth]{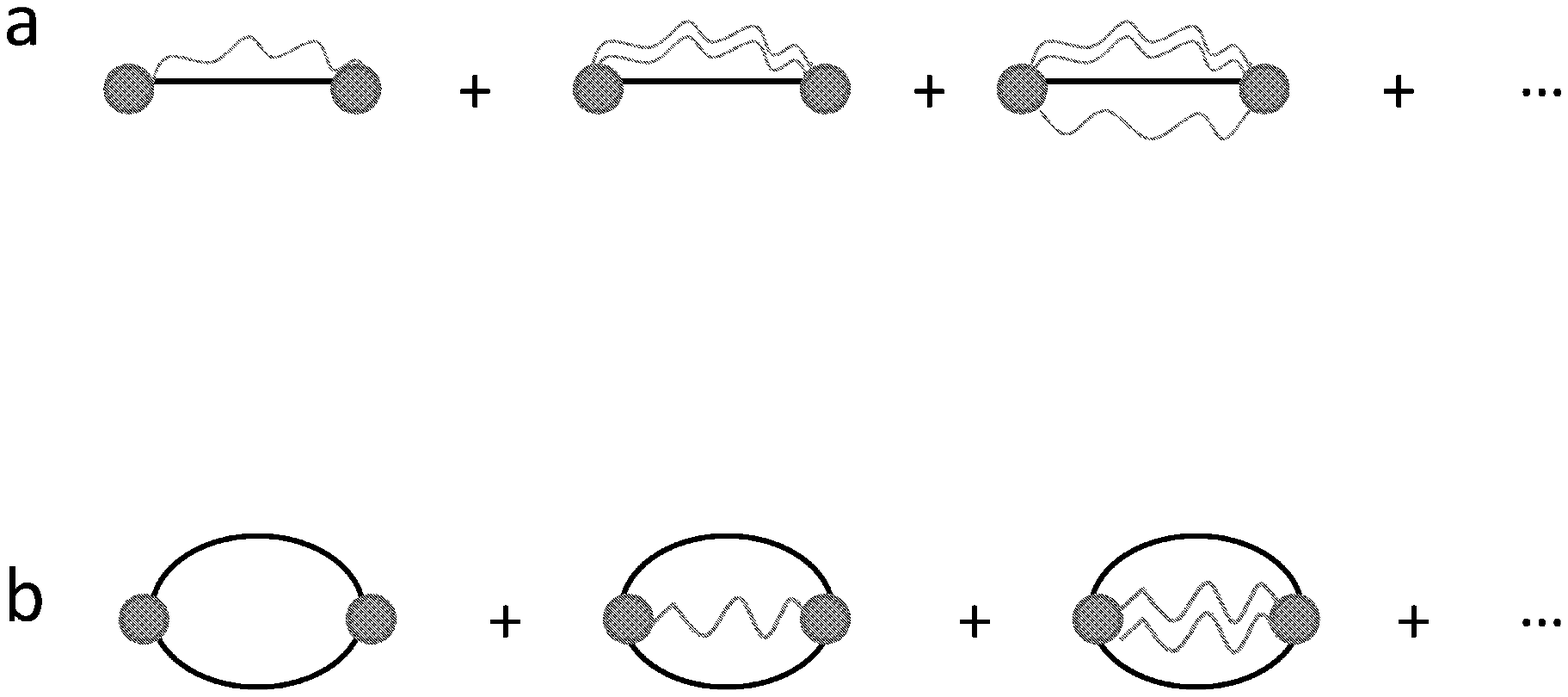} \vskip -0.5mm
\end{center}
\caption{(Color online) A few diagrams contributing to the second-order in $%
1/\protect\lambda$ polaron ($a$) and phonon ($b$) self-energies with
multi-phonon vertexes. Straight and wavy lines correspond to the polaron and
phonon propagators, respectively. }
\label{sigma}
\end{figure}

\section{Weak-coupling  regime}
For comparison, let us first consider the weak-coupling limit, where not
only $\lambda < 0.5$ but also the number of phonons dressing the carrier is
small, $E_p/\hbar \omega_0 \ll 1$. In this limit one can expand $\hat{X}_i$
in Eq.(\ref{trans}) in powers of $\gamma(q)$ keeping just single-phonon
transitions so that (in the momentum representation)
\begin{eqnarray}
\tilde{H}&\approx&\sum_{\mathbf{k},s}\xi_\mathbf{k }c_{\mathbf{k}%
,s}^{\dagger }c_{\mathbf{k},s}+H_{ph}+\cr && \sum_{\mathbf{q,k},s} \tilde{M}(%
\mathbf{k,q})c_{\mathbf{k+q},s}^{\dagger }c_{\mathbf{k},s}(d_\mathbf{q}-d_{-%
\mathbf{q}}^\dagger),  \label{transweek}
\end{eqnarray}
where $\xi_{\mathbf{k}}= E(\mathbf{k})-\tilde{\mu}$, $E(\mathbf{k})= -\sum_%
\mathbf{m} T(\mathbf{m}) \exp(i \mathbf{m}\cdot \mathbf{k})$ is the bare
band dispersion, and $\tilde{M}(\mathbf{k,q})=\gamma(q)[E(\mathbf{k+q})-E(%
\mathbf{k})]/\sqrt{2N}$ is the transformed EPI matrix element, renormalised
by the Coulomb repulsion. There are no other interactions left in the
transformed Hamiltonian since the bare Coulomb repulsion is nullified by the
Fr\"ohlich EPI.

Applying the BCS-Eliashberg formalism \cite{me} yields the master equation
for the superconducting order parameter, $\Delta(\omega_n,\mathbf{k})$,
\begin{equation}
\Delta(\omega_n,\mathbf{k})=k_BT \sum_{\mathbf{k}^\prime,\omega_{n^\prime}}{%
\frac{\tilde{M}(\mathbf{k},\mathbf{k}-\mathbf{k}^\prime)^2D(\omega_n-%
\omega_{n^\prime})\Delta(\omega_{n^\prime},\mathbf{k}^\prime) }{{%
\omega_{n^\prime}^2+\xi_{\mathbf{k}^\prime}^2+|\Delta(\omega_{n^\prime},%
\mathbf{k}^\prime)|^2}}},  \label{master}
\end{equation}
where $D(\omega_n-\omega_{n^\prime})=-\hbar
\omega_0/[(\omega_n-\omega_{n^\prime})^2+\hbar^2 \omega_0^2]$ is the phonon
propagator and $\omega_n=\pi k_BT (2n+1)$ are the Matsubara frequencies ($%
n=0,\pm 1,\pm 2,\pm 3,...$). Depending on a particular shape of the band
dispersion, Eq.(\ref{master}) allows for different symmetries of the order
parameter since EPI is not local \cite{aledwave}. Here we confine our
analysis to a simple estimate of T$_c$ by assuming a $\mathbf{k}$%
-independent gap function, $\Delta(\omega_n)$. Then factorizing the kernel
in Eq.(\ref{master}) on the "mass shell", $E(\mathbf{k}^\prime)-E(\mathbf{k}%
)=\omega_{n^\prime}-\omega_n$ and linearizing Eq.(\ref{master}) with respect
to the gap function one obtains the familiar estimate of the critical
temperature, $k_BT_c \approx \hbar \omega_0 \exp[ -1/(\lambda-\mu_c^\star)]$%
, where $\mu_c^\star=\lambda /(1+\lambda L)$ is the Coulomb
pseudopotential. In our case the weak-coupling BCS superconductivity
with $k_BT_c \ll \hbar \omega_0$ exists \emph{exclusively} due to
the "Tolmachev-Morel-Anderson" logarithm
$L=\ln(\tilde{\mu}/\hbar\omega_0) > 1$, if the EPI is retarded (i.e.
$\hbar \omega_0 < \tilde{\mu}$).

\section{Strong-coupling regime}
Actually the number of virtual phonons in the polaron cloud is large in
oxides and some other polar lattices, $E_p/\hbar \omega_0 \gg 1$  with the
characteristic (oxygen) optical phonon frequency $\hbar \omega_0 \lesssim 80$
meV, so that multi-phonon vertexes are essential in the expansion of the
hopping operator $\hat{\sigma}_{ij}$. To deal with this challenging problem
let us single out the coherent hopping in Eq.(\ref{trans}) averaging $\hat{%
\sigma}_{ij}$ with respect to the phonon vacuum, and consider the remaining
terms as perturbation, $\tilde{H}=H_{0}+H_{p-ph}$. Here
\begin{equation}
H_{0}=-\sum_{i,j}(t_{ij}\delta_{ss^\prime}+\tilde{\mu} \delta_{ij})c_{i}^{\dagger
}c_{j}+H_{ph}
\end{equation}
 describes free phonons and polarons coherently propagating in
a narrow band with the exponentially diminished hopping integral, $t_{ij}=T(%
\mathbf{m}-\mathbf{n})\exp[-g^2(\mathbf{m}-\mathbf{n})]$,
\begin{equation}
g^2(\mathbf{m})={\frac{1}{{2N}}}\sum_\mathbf{q}\gamma(q)^2 [1-\cos(\mathbf{q
}\cdot \mathbf{m})],  \label{t}
\end{equation}
and
\begin{equation}
H_{p-ph}=\sum_{i,j}(t_{ij}-\hat{\sigma}_{ij})\delta_{ss^\prime}c_{i}^{%
\dagger }c_{j}
\end{equation}
 is the residual polaron-multiphonon interaction, which is a
perturbation at large $\lambda$. In the diagrammatic technique the
corresponding vertexes have any number of phonon lines as shown in Fig.\ref%
{sigma} for the second-order in $H_{p-ph}$ polaron self-energy ($\Sigma_p
\approx - E_p/2z \lambda^2$) and the phonon self-energy ($\Sigma_{ph}\approx
-x\hbar\omega_0/z \lambda^2$) \cite{ale92}, where $z$ is the lattice
coordination number and $x$ is the atomic density of carriers. Hence the
perturbation expansion in $1/\lambda $ is applied if $\lambda \gg 1/\sqrt{2z}
$ \cite{fir2,ale92}. Importantly there is no structural instability in the
strong coupling regime since $|\Sigma_{ph}|\ll \hbar \omega_0$ \cite{ale92}.

The LF transformation, Eq.(\ref{trans}) is exact for any adiabatic ratio $%
\hbar \omega_0/T(a)$. However, if the perturbation expansion in $1/\lambda$
is restricted by lowest orders, then it significantly overestimates polaron
masses in the adiabatic regime, $\hbar \omega_0/T(a) < 1$, for the case of
the short-range (Holstein) EPI \cite{aledev} (here $T(a)$ is the
nearest-neighbor bare hopping integral). The polaronic band narrowing
factor, $\exp(-g^2)$ becomes very small for this EPI in the strong-coupling
regime, which would eliminate any possibility of high temperature
superconductivity and even metallicity of the small Hosltein polarons.

In our case of the long-range (Fr\"ohlich) EPI, Quantum Monte-Carlo
simulations \cite{alekor} show that the LF  transformation provides
numerically accurate polaron masses already in the zero order of the
inverse-coupling expansion both in the  adiabatic regime as well as in the
non-adiabatic one for \emph{any } strength of the Fr\"ohlich EPI. Moreover,
such small  polarons \cite{alekor} and  small
bipolarons \cite{ale96} are perfectly mobile in the relevant range of the
coupling and the adiabatic ratio \cite{aledev}.

The perturbation $H_{p-ph}$ has no diagonal matrix elements with
respect to phonon occupation numbers. Hence it can be removed from
the Hamiltonian in the first order using a second canonical
transformation ${\mathcal{H}} =e^{S_2}\tilde{H} e^{-S_2}$ with
\begin{equation}
(S_2)_{n^\prime n}=\sum_{i,j} {\langle n^\prime
|(\hat{\sigma}_{ij}-t_{ij})c_{i}^{\dagger }c_{j}|n\rangle
\over{E_{n^\prime}-E_n}}.
\end{equation}
Here $E_n, E_{n^\prime}$ and $%
|n\rangle, |n^\prime\rangle$ are the energy levels and the eigenstates of $%
H_0$, respectively. Taking into account that the polaron Fermi
energy is small compared with the phonon energy at strong coupling
and/or sufficiently
low doping \cite{alebra}, one can neglect its contribution to $%
E_{n^\prime}-E_n \approx \hbar \omega_0 \sum_{\mathbf{q}} n^\prime _{\mathbf{%
q}}$ and project the second-order in $1/\lambda$ Hamiltonian $\mathcal{H}$
onto the phonon vacuum $|0\rangle$ with the following result
\begin{eqnarray}
\mathcal{H}&=&-\sum_{i,j}(t_{ij}\delta_{ss^\prime}+\tilde{\mu}
\delta_{ij})c_{i}^{\dagger }c_{j}-\cr && \sum_{\mathbf{mn}\mathbf{m}^\prime
\mathbf{n}^\prime,ss^\prime} V_{\mathbf{m n}}^{\mathbf{m}^\prime \mathbf{n}%
^\prime}c_{\mathbf{m}s}^{\dagger}c_{\mathbf{n}s}c_{\mathbf{m}^\prime
s^\prime}^{\dagger}c_{\mathbf{n}^\prime s^\prime},
\end{eqnarray}
where
\begin{eqnarray}
&&V_{\mathbf{mn}}^{\mathbf{m}^\prime \mathbf{n}^\prime}=iT_{ij}T_{i^%
\prime j^\prime}\int_0^{\infty}dt e^{-\delta t}\times \cr
&&\langle 0|[\hat{X}%
^{\dagger}_i(t)\hat{X}_j(t)-e^{-g^2(\mathbf{m}-\mathbf{n})}]\hat{X}^{\dagger}_{i^\prime} \hat{X}%
_{j^\prime}|0\rangle, \label{int}
\end{eqnarray}
 and $\hat{X}^{\dagger}_i(t)$ is the Heisenberg
multi-phonon operator obtained by replacing $d_{q}$ in  $\hat{X}^{\dagger}_i$
with $d_{q}\exp (i \omega_0 t)$. Calculating the integral, Eq.(\ref{int})
 with $\delta
\rightarrow +0$ yields
\begin{equation}
V_{\mathbf{mn}}^{\mathbf{m}^\prime \mathbf{n}^\prime}={\frac{%
t_{ij}t_{i^\prime j^\prime}}{{\hbar \omega_0}}} \sum_{k=1}^\infty {\frac{f(%
\mathbf{mn},\mathbf{m}^\prime \mathbf{n}^\prime)^k}{{k! k}}},  \label{V2}
\end{equation}
where $f(\mathbf{mn},\mathbf{m}^\prime\mathbf{n}^\prime)=(1/2N) \sum_\mathbf{%
q}\gamma(q)^2 [\cos(\mathbf{q} \cdot (\mathbf{m}-\mathbf{n}^\prime))+\cos(%
\mathbf{q} \cdot (\mathbf{n}-\mathbf{m}^\prime))-\cos(\mathbf{q} \cdot (%
\mathbf{m}-\mathbf{m}^\prime))-\cos(\mathbf{q} \cdot (\mathbf{n}-\mathbf{n}%
^\prime))]$.

\section{Polaronic $t-J_p$ Hamiltonian}
All matrix elements, Eq.(\ref{V2}), of the polaron-polaron interaction are
small compared with the polaron kinetic energy except the \emph{exchange}
interaction, $J_p(\mathbf{m}-\mathbf{n}) \equiv V_{\mathbf{mn}}^{\mathbf{n}
\mathbf{m}}$ such that $f(\mathbf{mn},\mathbf{m}^\prime\mathbf{n}%
^\prime)=2g^2(\mathbf{m}-\mathbf{n})$. Using $\sum_{k=1}^\infty y^k/k! k=
-C-\ln(y)+Ei^\star(y)$ with $C\approx 0.577$ and $Ei^\star(y) \approx e^y/y$
(for large $y$) one obtains a substantial $J_p(\mathbf{m})=T^2(\mathbf{m}%
)/2g^2(\mathbf{m}) \hbar \omega_0 $, which is much larger than the polaron
hopping integral, $t/J_p \propto 2\hbar \omega_0 g^2 e^{-g^2}/T(a) \ll 1$ in
the strong coupling limit. Here $t$ is the nearest-neighbor polaron hopping
integrals. Keeping only this exchange we finally arrive with the polaronic
"t-J$_p$" Hamiltonian,
\begin{eqnarray}
\mathcal{H}&=&-\sum_{i,j}(t_{ij}\delta_{ss^\prime}+\tilde{\tilde{\mu}}
\delta_{ij})c_{i}^{\dagger }c_{j} \cr &+&2 \sum_{\mathbf{m} \neq \mathbf{n}}
J_p(\mathbf{m}-\mathbf{n}) \left(\vec{S}_\mathbf{m} \cdot \vec{S}_\mathbf{n}+%
{\frac{1}{{4}}}\hat{n}_\mathbf{m}\hat{n}_\mathbf{n}\right),  \label{tJ}
\end{eqnarray}
where $\vec{S}_\mathbf{m}=(1/2)\sum_{s,s^\prime}c^\dagger_{\mathbf{m}s}\vec{%
\tau}_{ss^\prime} c_{\mathbf{m}s^\prime}$ is the spin 1/2 operator ($\vec{%
\tau}$ are the Pauli matrices), $\hat{n}_\mathbf{m}=\sum_s \hat{n}_i$, and $%
\tilde{\tilde{\mu}}=\tilde{\mu}+\sum_{\mathbf{m}}J_p(\mathbf{m})$ is the
chemical potential further renormalized by $H_{p-ph}$.

There is a striking difference between this polaronic t-J$_p$ Hamiltonian
and the familiar t-J model derived from the repulsive Hubbard U Hamiltonian
in the limit $U\gg t$ omitting the so-called three-site hoppings and EPI
\cite{tJ}. The latter model acts in a projected Hilbert space constrained to
no double occupancy. Within this standard t-J model the bare transfer
amplitude of electrons ($t$) sets the energy scale for incoherent transport,
while the Heisenberg interaction ($J \propto t^2/U$) allows for spin flips
leading to coherent hole motion with an effective bandwidth determined by $J
\ll t$. Using the Gutzwiller-type approximation to remove the constraint
results in an unconstrained t-J model also containing a band narrowing, but
purely electronic rather than phononic origin \cite{rice}. On the contrary
in our polaronic t-J$_p$ Hamiltonian, Eq.(\ref{tJ}) there is no constraint
on the double on-site occupancy since the Coulomb repulsion is negated by
the Fr\"ohlich EPI. The polaronic hopping integral $t$ leads
to the coherent (bi)polaron band and the antiferromagnetic exchange of purely phononic
origin $J_p$  bounds polarons into small superlight inter-site bipolarons.
Last but not least the difference is in the "+" sign in
the last term of Eq.(\ref{tJ}) proportional to $\hat{n}_\mathbf{m}\hat{n}_%
\mathbf{n}$, which protects the ground superconducting state from the
bipolaron clustering, in contrast with the "-" sign in
the similar term of the standard t-J model, where
the phase separation is expected at sufficiently large J \cite{kiv}.

\begin{figure}[tbp]
\begin{center}
\includegraphics[angle=-00,width=0.47\textwidth]{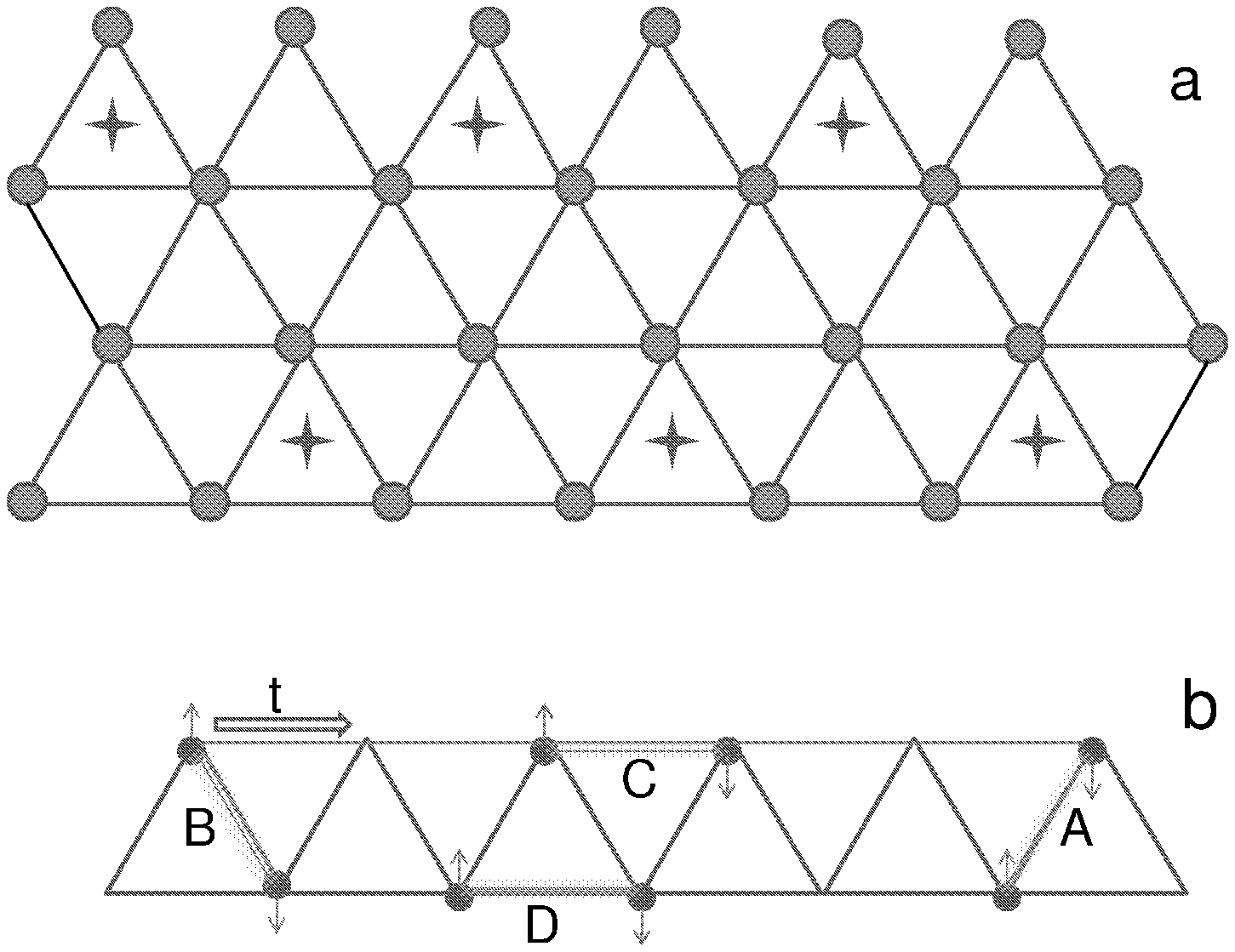} \vskip -0.5mm
\end{center}
\caption{(Color online) A plane of the 3D polar lattice of anions
(circles) and cations (crosses) (a) with doped carriers on anions
bound by the polaronic exchange into four degenerate singlet
bipolarons $A,B,C$ and $D$ (b). } \label{toy}
\end{figure}

The cancelation of the bare Coulomb repulsion by the Fr\"ohlich EPI
is accurate up to  a  $1/\epsilon_0$ correction. This correction
produces a long-range residual repulsion of (bi)polarons in the
transformed Hamiltonian, Eq.(\ref{tJ}), which is small as soon as
$\epsilon_0 \gg e^2/a J_p$. The residual repulsion results in some
screening of the Coulomb interactions  responsible for the doping
dependence of the bipolaron binding energy and of the (bi)polaron
mass \cite{alebra}. In  layered polar insulators the static
dielectric constant  could be anisotropic, which together with local
field corrections might result in different EPI matrix elements for
 in-plane and out-of-plane polarised optical phonons, respectively. The difference is not
considered here.

\section{Projection onto bipolaronic Hamiltonian and high T$_c$}
The polaronic t-J$_p$ Hamiltonian, Eq.(\ref{tJ}) is analytically solvable in
the limit of sufficiently low atomic density of carriers, $x\ll 1$.
Neglecting the first term in $\mathcal{H}$, which is the polaron kinetic
energy proportional to $t \ll J_p$, one can readily diagonalise the
remaining spin-exchange part of the Hamiltonian. Its ground state is an
ensemble of inter-site singlet bipolarons with the binding energy $%
\Delta_b=J_p$ localised on nearest neighbor sites. Such small bipolarons
repel each other and single polarons via a short-range repulsion of about $%
J_p$.

The kinetic energy operator in Eq.(\ref{tJ}) connects singlet
configurations in the first and higher orders with respect to the polaronic
hopping integrals. Taking into account only the lowest-energy degenerate
singlet configurations and discarding all other configurations one can
project the t-J$_p$ Hamiltonian onto the inter-site bipolaronic Hamiltonian
using the bipolaron annihilation operators $B_{\mathbf{m}}=2^{-1/2} (c_{%
\mathbf{m}\uparrow}c_{\mathbf{m}+\mathbf{a}\downarrow}-c_{\mathbf{m}%
\downarrow}c_{\mathbf{m}+\mathbf{a}\uparrow})$, where $\mathbf{a}$ connects
nearest neighbors \cite{ale96}. These operators are similar to  the bond-order operators
introduced later by Newns and Tsuei \cite{newns}, which are weakly coupled in their model with
the single-plane lattice vibrations via a nonlinear (two-phonon) EPI [Eq.(9)
in Ref. \cite{newns}]. Actually it is well known that the nonlinear anharmonic corrections to EPI
are small compared with the linear Fr\"ohlich interaction in real three-dimensional solids,
which makes the two-dimensional model of Ref.\cite{newns} of cuprate superconductors unfeasible.

 Our strong-coupling projection is
illustrated using a polar lattice, sketched in Fig.\ref{toy}a,  of
anion-cation triangular planes (the in-plane lattice constant is $a$ and the
nearest-neighbor hopping distance is $a/2$) separated by the out-of-plane
lattice constant $c$. For a zig-zag ladder-fragment of the lattice, Fig.\ref%
{toy}b, the projected bipolaronic Hamiltonian in the nearest-neighbor
hopping approximation is
\begin{eqnarray}
H_b &=& -t \sum_n B_n^\dagger A_n+D_n^\dagger B_n+D_n^\dagger A_n
+C_n^\dagger B_n + A_{n+1}^\dagger C_n \cr &+& A_{n+1}^\dagger
B_n+B_{n-1}^\dagger A_n +C_{n-1}^\dagger A_n +H.c.,  \label{bip}
\end{eqnarray}
where $A,B,C,D$ are annihilation operators of the four degenerate singlets.
Fourier transformation $H_b$ yields four bipolaronic bands, $%
E_{1,2}(K)=-t[\cos(Ka/4)\pm \sqrt{1+4 \sin(Ka/8)^4}]$, $E_{3,4}(K)=t[%
\cos(Ka/4)\pm \sqrt{1+4 \cos(Ka/8)^4}]$ with the center-of-mass momentum $%
\hbar K$. Expanding in powers of $K$ one obtains the effective mass
of these small singlets, $m^{**}=10 m^*$, where  $m^*=2\hbar^2/5
t(a/2)^2$ is the polaron mass. A similar Hamiltonian can be derived
 also for a square lattice, if
next-nearest-neighbor hopping integrals are taking into account
\cite{alekor2}.

Small bipolarons are hard-core bosons with the short-range repulsion
of the radius $r=a/2$ and a huge anisotropy of their effective mass
since their inter-plane hopping is possible only in the second order
of $t$ \cite{alekabmot}. The occurrence of superconductivity in
bipolaronic strong coupling systems   is not controlled by the
pairing strength, but by the phase coherence among the electron
pairs \cite{aleran}. While in two dimensions Bose condensation does
not occur in either the ideal or the interacting system, there is a
phase transition to a superfluid state at
\begin{equation} T_c={2\pi
n_b \hbar^2\over{k_B m^{**} \ln[\ln(1/n_b r^2)]}}
\end{equation}
 in
the dilute Bose gas \cite{popov,fisher} (here $n_b$ is the boson
density per unit area). Using Eqs.(\ref{shift}, \ref{t}) we obtain
$E_p \approx 0.4 E_c$ and $g^2 \approx 0.18 E_p/\hbar \omega_0$,
allowing for a quantitative estimate of T$_c$
(here $E_c=2 e^2 c/\pi \kappa a^2$). With typical values of $a=0.4$ nm, $%
c=1.2$nm, $\kappa=5$,  the bare band mass $m=m_e$, $\hbar \omega_0=80$ meV
and the moderate atomic density of polarons, $x=0.1$ (avoiding an overlap of
bipolarons) one obtains $E_p \approx 0.55$ eV, $g^2\approx 1.24$, and $T_c
\approx 205$K. Importantly, the projection procedure of reducing Eq.(\ref{tJ}%
) to Eq.(\ref{bip}) is well justified since the ratio $t/ J_p
\approx 0.1$ is small and $k_BT_c \ll J_p$, so that only the lowest
singlet configurations can be included while discarding the others.

In conclusion, it seems very likely that a peculiar cancelation of
the long-range Coulomb repulsion by the long-range Fr\"ohlich EPI
can help much in producing high-temperature superconductivity in
doped polar insulators such as cuprates and other oxides, for
instance BaKBiO. The polaronic t-J$_p$ Hamiltonian, Eq.(\ref{tJ})
derived here from the bare long-range Coulomb interactions could
provide a novel avenue for analytical and computational studies of
superconductivity in complex ionic lattices since the repulsive
Hubbard U model and its strong-coupling t-J projection  do not
explain high T$_c$ \cite{alekab2011}.

\section*{Acknowledgements}
The author thanks Alexander Bratkovsky, Janez Bon\v{c}a, Jozef
Devreese, Holger Fehske, Viktor Kabanov, Dragan Mihailovi\'{c},
Peter Prelovsek, John Samson  and G. Sica for helpful discussions.
This work was partially supported by the Royal Society (London).


\begin{thebibliography}{99}
\bibitem{aledev}  Alexandrov A. S. and  Devreese J. T., \emph{Advances in
Polaron Physics }(Springer, Berlin 2009).

\bibitem{alebra}  Alexandrov A. S. and  Bratkovsky A. M., Phys. Rev. Lett.
\textbf{105}, 226408 (2010).

\bibitem{me}  Eliashberg G. M., Zh. Eksp. Teor. Fiz. \textbf{39}, 1437 (1960)
[Sov. Phys.--JETP  \textbf{12}, 1000 (1960)].

\bibitem{ale0}  Alexandrov A. S., Zh. Fiz. Khim. \textbf{57}, 273 (1983)
[Russ. J. Phys. Chem. \textbf{57}, 167 (1983)].

\bibitem{LDB1977}  Lemmens L. F.,  Brosens F., and  Devreese J. T., Physica
Status Solidi (b) \textbf{82}, 439 (1977).

\bibitem{aleran}  Alexandrov A. S. and  Ranninger J., Phys. Rev. B \textbf{23}%
, 1796 (1981); Phys. Rev. B \textbf{24}, 1164 (1981).

\bibitem{ale96}  Alexandrov A. S., Physica C (Amsterdam) \textbf{ 182}, 327 (1991); Phys. Rev. B \textbf{53}, 2863 (1996).
\bibitem{inter}
 Aubry S., J. Physique (France) IV Colloque C\textbf{2}, 349 (1993);
Bon\v{c}a J. and  Trugman S. A., Phys. Rev. B \textbf{64}, 094507 (2001));
Alexandrov A. S. and  Kornilovitch P. E., J. Phys.: Condens. Matter \textbf{14},
5337 (2002);  Macridin A.,  Sawatzky G. A. and  Jarrell M., Phys. Rev. B\textbf{%
\ 69}, 245111 (2004):  Hague J. P.,  Kornilovitch P. E.,  Samson J., and  Alexandrov A. S., Phys. Rev. Lett. \textbf{98},
037002 (2007).

\bibitem{trugman} Fehske  H. and  Trugman S. A., in \emph{Polarons in Advanced
Materials}, ed. A. S. Alexandrov (Springer, Dordrecht 2007) pp 393-461;
 Mishchenko A. S. and  Nagaosa N., ibid, pp 503-544.

\bibitem{bonca}  Vidmar L.,  Bonca J.,  Maekawa S., and  Tohyama T., Phys. Rev. Lett. \textbf{103}
186401 (2009).

\bibitem{bauer}  Bauer T. and  Falter C., Phys. Rev. B \textbf{80}, 094525
(2009).

\bibitem{mahan} Mahan  G. D., \textit{Many-Particle Physics} (Plenum, New
York 1990).

\bibitem{fir}  Lang I.~G. and  Firsov Y.~A., Zh. Eksp. Teor. Fiz. \textbf{43},
1843 (1962) [Sov. Phys. JETP \textbf{16}, 1301 (1962)].

\bibitem{aledwave}  Alexandrov A. S., Phys. Rev. B \textbf{77}, 094502 (2008).

\bibitem{ale92}  Alexandrov A. S., Phys. Rev. B \textbf{46}, 2838 (1992).

\bibitem{fir2}  Eagles D. M., Phys. Rev. \textbf{145}, 645 (1966);
 Gogolin A. A., Phys. Status Solidi B \textbf{109}, 95 (1982).

\bibitem{alekor}  Alexandrov A. S. and  Kornilovitch P. E., Phys. Rev. Lett.
\textbf{82}, 807 (1999).

\bibitem{tJ} Hirsch J. E., Phys. Rev. Lett. \textbf{54}, 1317 (1985);  Spalek J., Phys. Rev B \textbf{37}, 533 (1988);
 Gros C.,  Joynt R. and Rice  T. M., Phys. Rev. B \textbf{36}, 381 (1987).

\bibitem{rice}  Zhang F. C.,  Gros C.,  Rice T. M., and  Shiba H., Supercond.
Sci. Technol. \textbf{1}, 36 (1988).

\bibitem{kiv}  Emery V. J.,  Kivelson S. A. and  Lin H. Q., Phys. Rev. Lett. \textbf{64}, 475 (1990).

\bibitem{newns}  Newns D. M. and  Tsuei C. C., Nature Physics \textbf{3},
184 (2007).

\bibitem{alekor2} Alexandrov A. S. and  Kornilovitch P. E., J. Phys.: Condens. Matter
\textbf{14},
5337 (2002).
\bibitem{alekabmot}  Alexandrov A. S.,  Kabanov V. V. and Mott  N. F., Phys. Rev. Lett. \textbf{77}, 4796 (1996)
\bibitem{popov}  Popov N., Theor. Math. Phys. \textbf{11}, 565 (1972).

\bibitem{fisher}  Fisher D. S. and  Hohenberg P. C., Phys. Rev. B \textbf{37},
4936 (1988).

\bibitem{alekab2011} Alexandrov A. S. and Kabanov V. V., Phys. Rev. Lett. \textbf{106}, 136403 (2011).










\end{thebibliography}
\end{document}